\documentclass[12pt]{iopart}
\usepackage{graphicx}
\begin{document}

\title{Growth and form of melanoma cell colonies}

\author{Massimiliano Maria Baraldi$^1$, Alexander A. Alemi$^2$, James P. Sethna$^2$, Sergio Caracciolo$^3$, Caterina A. M. La Porta$^4$ 
and Stefano Zapperi $^{5,6}$}
\address{$^1$ Physics Department, University of Milano, via Celoria 16, 20133 Milano, Italy}
\address{$^2$ LASSP, Department of Physics, Clark Hall, Cornell University, Ithaca, NY 14853-2501, USA}
\address{$^3$ Physics Department, University of Milano and INFN, via Celoria 16, 20133 Milano, Italy }
\address{$^4$ Department of Biosciences, University of Milano, via Celoria 26, 20133 Milano, Italy}
\address{$^5$ CNR-IENI, Via R. Cozzi 53, 20125 Milano, Italy}
\address{$^6$  ISI Foundation,Via Alassio 11/C, 10126 Torino, Italy}
\ead{stefano.zapperi@cnr.it}
\begin{abstract}
We study the statistical properties of melanoma cells colonies grown in vitro analyzing the 
results of crystal violet assays at different concentrations of initial plated cells and for different
growth times. The distributions of colony sizes is well described by a continuous time branching process.
To characterize the shape fluctuations of the colonies, we compute the distribution of eccentricities.
The experimental results are compared with numerical results for models of random division of elastic
cells showing that experimental results are best reproduced by restricting cell division to the outer
rim of the colony. Our results serve to illustrate the wealth of information that can be extracted by
a standard experimental method such as the crystal violet assay.
\end{abstract}

\section{Introduction}

Understanding how tumors grow is a long-standing problem whose complexity stems from the 
interplay of several factors including cancer cell heterogeneity, interactions with the environment 
and stochastic mutations during progression. Some interesting features of tumor growth can, however,
be inferred from in vitro experiments where cancer cells duplication and spreading can be observed
in ideal conditions. In particular, colony formation assays have a practical interest in drug testing where
one compares cell colonies formed with or without a drug.  Due to its relative simplicity, cancer colony formation has
attracted the interest of statistical physicists \cite{kimmel1991,bru2003,buceta2005,huergo2011,huergo2012} who mostly
focused on the morphological properties of individual colonies - e.g. the colony boundary, which is well described by self-affine scaling \cite{bru2003,buceta2005,huergo2011,huergo2012}. Colony formation also represents the ideal playground to test mathematical models for tumor growth based on individual cell mechanics \cite{drasdo1995,galle2005,galle2009,drasdo2012} ,  cellular automata \cite{loeffler1987,block2007} or differential equations \cite{chatelain2011}. 

In typical cancer cell colony growth experiments, such as in the crystal violet assay, biologists
study the distribution of colonies formed in multiwells. In this case, one is usually interested just in 
the number of colony forming clones. The experiments, however, provide a wealth of other useful
information concerning the distributions of colony sizes and shapes that are rarely analyzed in detail.
In this paper, we illustrate this point by studying the statistical properties of colonies formed by melanoma cells in vitro
for different growth times. In particular, we consider the distribution of sizes and shapes. By using automatic
image analysis methods, we are able to study with relative ease thousands of colonies formed by 
hundreds of cells. As a comparison, earlier studies based on manual counting of the cells in the colonies could perform
statistics over around 50 colonies \cite{kimmel1991}. Here we show that  the experimentally measured colony size distributions are quantitatively described by a branching process \cite{harris89,kimmel2002}, a class of models that have been used extensively in the past to model growth of  stem cells \cite{vogel1968,matioli1970,potten1988,clayton2007,antal2010,itzkovitz2012} and cancer cells \cite{kimmel1991,michor2005,ashkenazi2008,michor2008,tomasetti2010,laporta2012}.

The main limitation of branching processes is due to their mean-field nature that does not take into account 
the geometry of the cellular arrangement inside a tumor.  Hence to understand the distribution of colony shapes,
quantified by their eccentricity, we use an individual cell models in which cells interact either by elastic interactions \cite{drasdo1995,galle2005}  or by simple geometrical hinderance, corresponding in the simplest example to the Eden growth model \cite{eden1961}. We find that the Eden model describes the experimental eccentricity data more accurately than the elastic interaction model.  Our results show that  a comparison between experimental and theoretical results provides useful indications for  building realistic models for cancer growth.

The paper is organized as follows: in section 2 we discuss the experiments and the data analysis, in section 3 and 4 we 
report the discuss size and shape distributions, respectively, and section 4 is devoted to conclusions.

\section{Cancer cell colony formation: experiments and data analysis}

\subsection{Colony growth}
To study the formation of melanoma cell colonies in vitro, we use human IGR39 cells, obtained from Deutsche Sammlung von Mikroorganismen und Zellkulturen GmbH and cultured as previously described \cite{taghizadeh2010}. IGR39 cells are derived from a primary amelanotic cutaneous tumor.
Cells are plated on 6 multiwell and stained after 8 days or 10 days. Next, cells are fixed with 3.7\% paraformaldeide (PFA) for 5 minutes and then stained for 30min with 0.05\% crystal violet solution. After two washing with tap water, the plates are drained by inversion for a couple of minutes.
In order to control the merging of different colonies, the experiments are performed with different initial cell concentrations
$N_0$. In particular, we use  $N_0=1,10,20,30,80,100,150,250$ cells/well for cells growing 8 days, and $N_0=1,10,50,100,150$ for
cells growing 10 days.

\subsection{Image analysis}
\begin{figure}[t]
\centering
\includegraphics[width=0.8\textwidth]{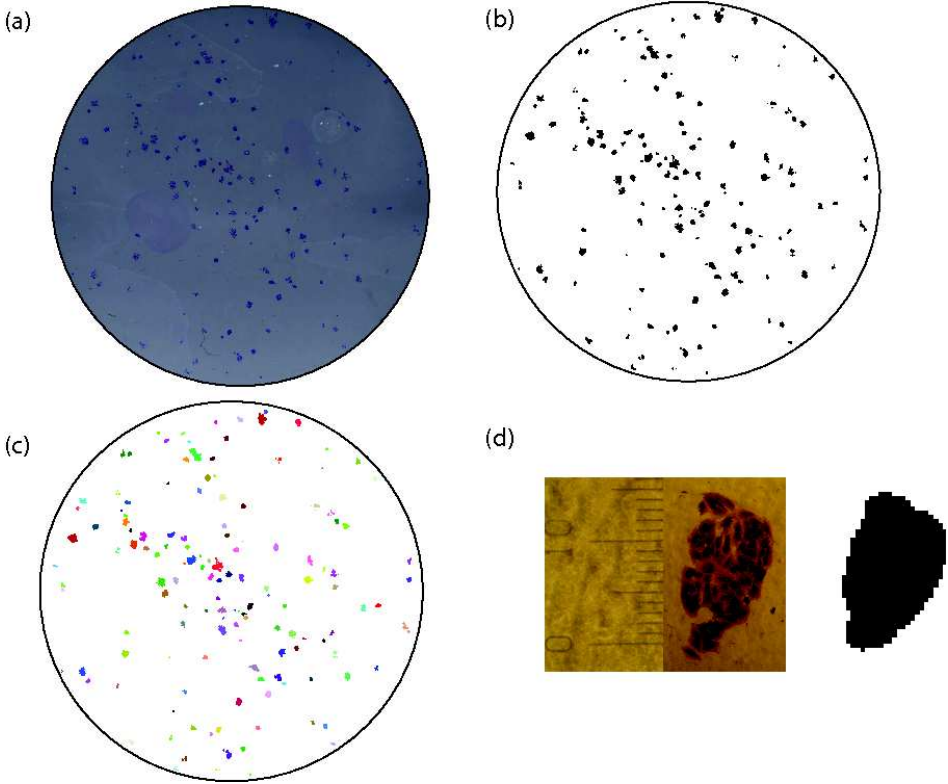}
\caption{An illustration of the cluster algorithm. a) Melanoma cell colonies formed after 8 days in a single
well. b)  The picture in transformed into a  black \& white image. c) Individual colonies, identified using the cluster
algorithm,  are colored here with random colors for visualization purposes. 
d) A digitized cluster can  be compared with an image of the same cluster obtained with a microscope in order to 
quantify the cell density versus area.}
\label{fig:1}
\end{figure}

We acquired images of the six wells by a simple scanner with resolution of 600 x 2400 dpi.  The color images are then transformed
into black and white  (see Fig. \ref{fig:1}). In order to do this automatically, avoiding  errors associated with shadows
and noise, we separate the colonies from the background using an edge detect difference of Gaussians algorithm \cite{marr1980}. 
In this way, even small colonies can be identified from the background and the transformation into a black and white image is insensitive
to the threshold used.  Some of the wells have to be discarded either because a considerable fraction of the cells detached or because the staining was too noisy to process. In addition, we avoid errors associated with the boundary of the well by considering only colonies inside an inner circle of radius $R=650$px, while the well had a radius of 800px. 

Next we perform a cluster analysis to isolate individual colonies. This is done in two steps: we first perform an Hoshen-Kopelman algorithm \cite{hoshen76} to identify clusters that are separated by at least one pixel. This results sometimes in very small clusters surrounding large ones. We think that this is due to the fact that individual cells can sometime separate from the boundary of the colony during the division
process.  We therefore devise an algorithm to join small clusters to big ones. We first divide small and large clusters depending on a threshold $S^*$ that we set equal to 100px. We then scan the lattice and if a small cluster is within a radius $r=8$px from a large cluster, we join them together into a single cluster.  We test the effect of the value of $S^*$ on the final colony size distribution and find only small variations
in the final outcome. Finally we convert the cluster size from pixels to cells  using a microscope endowed with the resolution of 1 $\mu$m. We count the number of cells contained in a set of colonies from which we  estimate a conversion factor 
$p =  0.14 \pm 0.05$ cells/px  or equivalently a cell corresponds to seven to eight px. 

\subsection{Cluster arrangement}
\begin{figure}[h]
\centering
\vspace{1.5cm}

\includegraphics[width=0.8\textwidth]{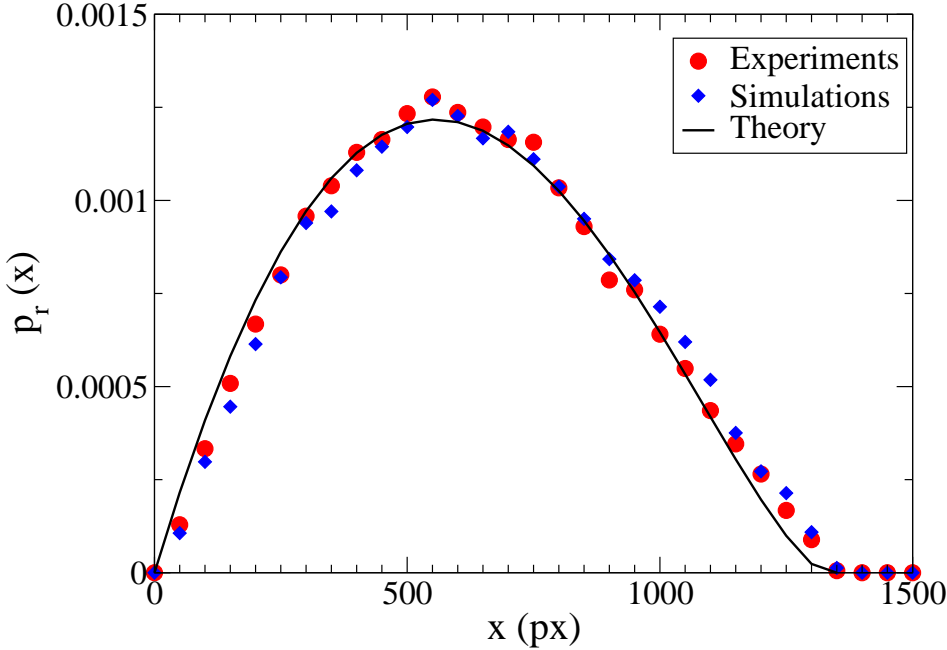}
\caption{The distribution of the distances between the colonies' center of mass compared with anlytical and
numerical predictions from a random Poisson process.}
\label{fig:CM}
\end{figure}

According to the experimental protocol, cells are initially spread randomly and uniformly on the wells. We check if this assumption is correct
by comparing the spatial arrangement of the colonies to a random process. 
This is achieved comparing the distribution of  the (Euclidean) distances between the center of mass of the colonies with the distribution $p_r (x) = p_r (x_{i,j} = x)$ of distances $x_{i,j}=|| \mathbf{x_i} - \mathbf{x_j} ||$ obtained from a Poisson process in a circle of radius $r$.  
Figure \ref{fig:CM} shows the experimental distribution of $p_r(x)$ and the simulated distribution compared with the analytical expression given by
\begin{equation}
p_r(x) = \frac{2 x}{r^2} \left( \frac{2}{\pi} \arccos \left(\frac{x}{2 r}\right) -\frac{x}{\pi r} \sqrt{1-\frac{x^2}{4 r^2}} \right).
\label{bareq}
\end{equation}
We also report the result of a numerical distribution obtained throwing random points in a circular well. The good agreement between experimental data and analytical curves shows that colonies are randomly distributed in the well without noticeable interactions between them.
In principle, cell motility could affect the final position of the colony. Time lapse microscopy shows, however, that in present experimental conditions cells move very little as compared with the typical distance between the colonies.

\subsection{Density dependence}
\begin{figure}[t]
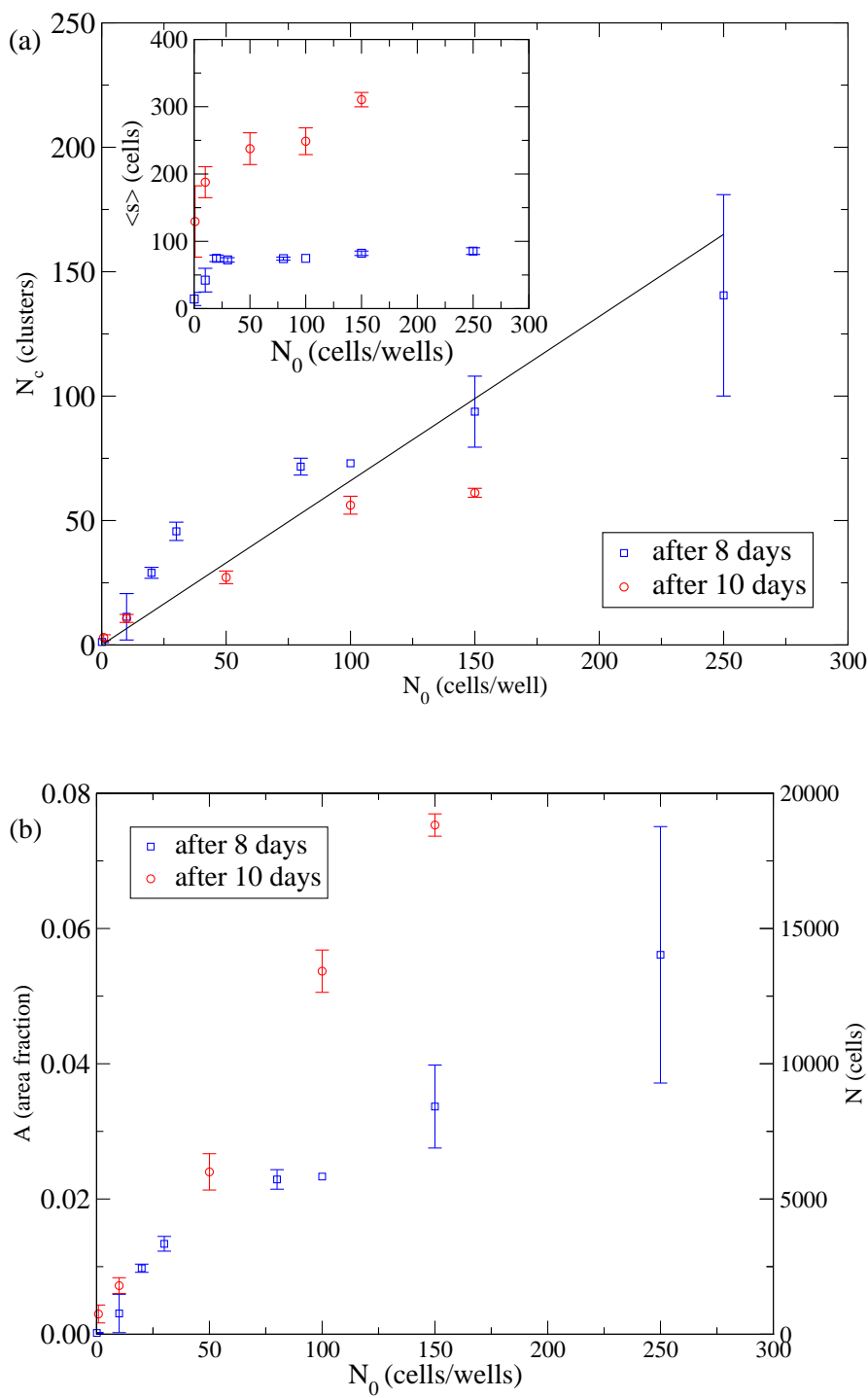

\centering
\includegraphics[width=0.75\textwidth]{fig3a.eps}

\vspace{1cm}

\includegraphics[width=0.75\textwidth]{fig3b.eps}
\caption{The role of the initial cell density in cluster statistics. a) The total number of colonies grows linearly with the number of the initial
cells $N_0$ present in each well but is almost independent of the incubation time. The slope of the line is the ratio of the area analyzed to the total area of the well. The inset shows that the average colony size is reasonably constant as the initial density is varied.
b) The fraction of area covered $A$ and the total number of cells $N$ grow linearly with the initial number of cells where the slope depends on the
incubation time. Error bars are the standard errors obtained considering data from different wells (typically six or fewer wells are considered).
}
\label{fig:cluster_number}
\end{figure}

Experiments are performed at different initial densities $N_0$ in order to test the dependence of the
resulting colonies on the initial condition. In Fig. \ref{fig:cluster_number}a we plot the number of colonies
as a function of $N_0$. As expected, the number of colonies $N_c$ is roughly equal to the number of cells plated
initially, providing a test of the validity of the cluster algorithm. Next in the inset of Fig. \ref{fig:cluster_number}a, we report
the average colony size $\langle s \rangle$ as a function of $N_0$. We  showing a little dependence, especially for colonies grown for 8 days.
This result provides indirect evidence that the merging of different colonies is a marginal effect. We expect that using
larger initial density would make the results less reliable. In  Fig. \ref{fig:cluster_number}
we report the fraction of total area $A$ that is covered by cells in each well as a function of the initial cell density $N_0$. 
Fig. \ref{fig:cluster_number} also reports the total number of cell in each well $N$ 
obtained by counting the number of occupied pixels and then dividing by the conversion factor $p=0.14$cells/px.
 As expected, both $A$ and $N$  grow linearly with $N_0$.

\section{Colony size distributions}
\begin{figure}[t]
\centering
\vspace{2cm}

\noindent\includegraphics[width=0.8\textwidth]{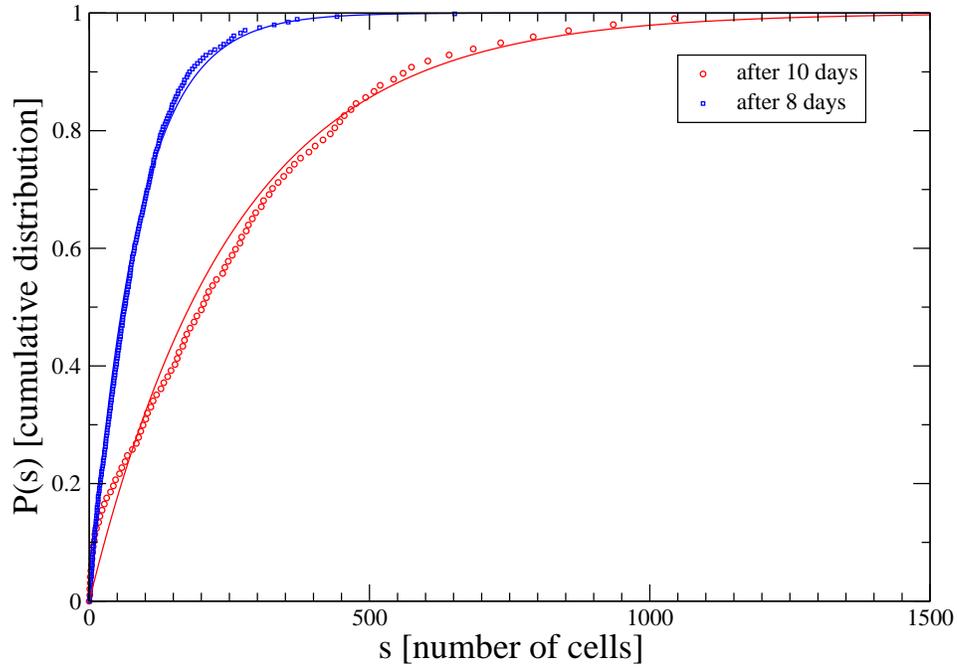}
\caption{The distribution of colony sizes after 8 and 10 days of growth. The distribution is obtained collecting
all the colonies obtained for all the different values of $N_0$, corresponding to 32 and 30 wells, 2347 and 1067 colonies
of average size $\langle S \rangle \simeq 85$ and  $\langle S \rangle \simeq 258$
for 8 and 10 days respectively. The curves represent the simulation of the continuum time branching process using the parameters obtained from the maximum likelihood estimate.}
\label{fig:cdf}
\end{figure}

In order to describe the fluctuations in the growth of melanoma cell colonies, we compute the cumulative distribution of 
colony sizes grown for 8 or 10 days, collecting together data obtained in different wells and for different 
values of $N_0$.  The cumulative distribution $P(s)$ is related to the  probabilty density function $p(s)$ by
\begin{equation}
P(s) = \sum_{s'=1}^{s} p(s').
\end{equation}
As shown in Fig \ref{fig:cdf}, colonies become larger as time passes and consequently the cumulative
distribution shifts to the right.

To understand more quantitatively the experimentally measured colony size distribution, we consider 
a simple continuous time branching process in which each cell duplicates at rate $\gamma$ and 
dies at rate $\beta$.  Here we are interested in the evolution of the size distribution of colony sizes $p(s,t)$, defined as
the probability to find a colony of size $s$ at time $t$, where time is measured in days.  
The probability density function evolves according to the following master equation
\begin{equation}
\frac{d p(s,t)}{dt} = \gamma (s-1) \; p(s-1,t) +\beta (s+1) \; p(s+1,t)-(\gamma+\beta) s \; p(s,t),
\label{eq:master_equation}
\end{equation}
starting with an initial condition $p(s,0)= \delta_{s,1}$.
From  Eq. \ref{eq:master_equation}, we can obtain an equation for the first moment, the average
colony size $\langle s \rangle$,
\begin{equation}
\frac{d\langle s \rangle }{dt} = (\gamma-\beta) \langle s \rangle,
\end{equation}
yielding an exponential growth
\begin{equation}
\langle s (t)\rangle = \exp [(\gamma-\beta)t].
\end{equation}

The model described by Eq. \ref{eq:master_equation} is a particular case of the {\it birth and death process}
and its explicit solution is given by (see Ref. \cite{harris89} page 104):
\begin{eqnarray}
p(0,t)&=&1 - \frac{\langle s (t)\rangle(\gamma-\beta)}{\gamma\langle s (t)\rangle-\beta } \\
p(s,t) &=&\langle s (t)\rangle \left( \frac{\gamma-\beta}{\gamma\langle s (t)\rangle-\beta }\right)^2
 \left(1- \frac{\gamma-\beta}{\gamma\langle s (t)\rangle-\beta }\right)^{s-1} ~~\mbox{ for } s \geq 1.
\end{eqnarray}
To obtain a quantitative comparison between theory and experiments, we 
employ the maximum likelihood method and estimate the best values for $\gamma$ and $\beta$,
with the constraints $\beta\geq 0$ and $\gamma\geq 0$.
In practice, using an iterative optimization scheme we find the values of $\gamma$ and 
$\beta$ that maximize the cost function given by
\begin{equation}
{\mathcal L} = \sum_i \log ( p(s_i^{(t=8)},8))+\sum_j \log p(s_j^{(t=10)},10)
\end{equation}
where $s_i^{(t=8)}$ and $s_j^{(t=10)}$ are the experimentally measured colony sizes
after 8 and 10 days, respectively. Using this scheme, the best fit yields $\gamma=0.55 $ divisions/day
and sets $\beta$  to its limiting value of zero. In the limit $\beta\to 0$, the model is equivalent to
the Yule problem \cite{harris89} and the solution is given in simpler form by \cite{harris89}
\begin{equation}
p(s,t) =e^{-\gamma t} (1-e^{-\gamma t})^{s-1} ~~\mbox{ for } s \geq 1.
\end{equation}
The fitted value for $\gamma$ can be compared with previous experiments on melanoma growth, but at
much higher cell density, which yielded $\gamma \simeq 0.4$ divisions/day \cite{laporta2012}. This is
compatible with the general idea that the growth rate decreases when the cell density is higher.

\section{Colony shapes}

\subsection{Eccentricities}
\begin{figure}[t]
\centering
\vspace{1cm}

\noindent\includegraphics[width=\textwidth]{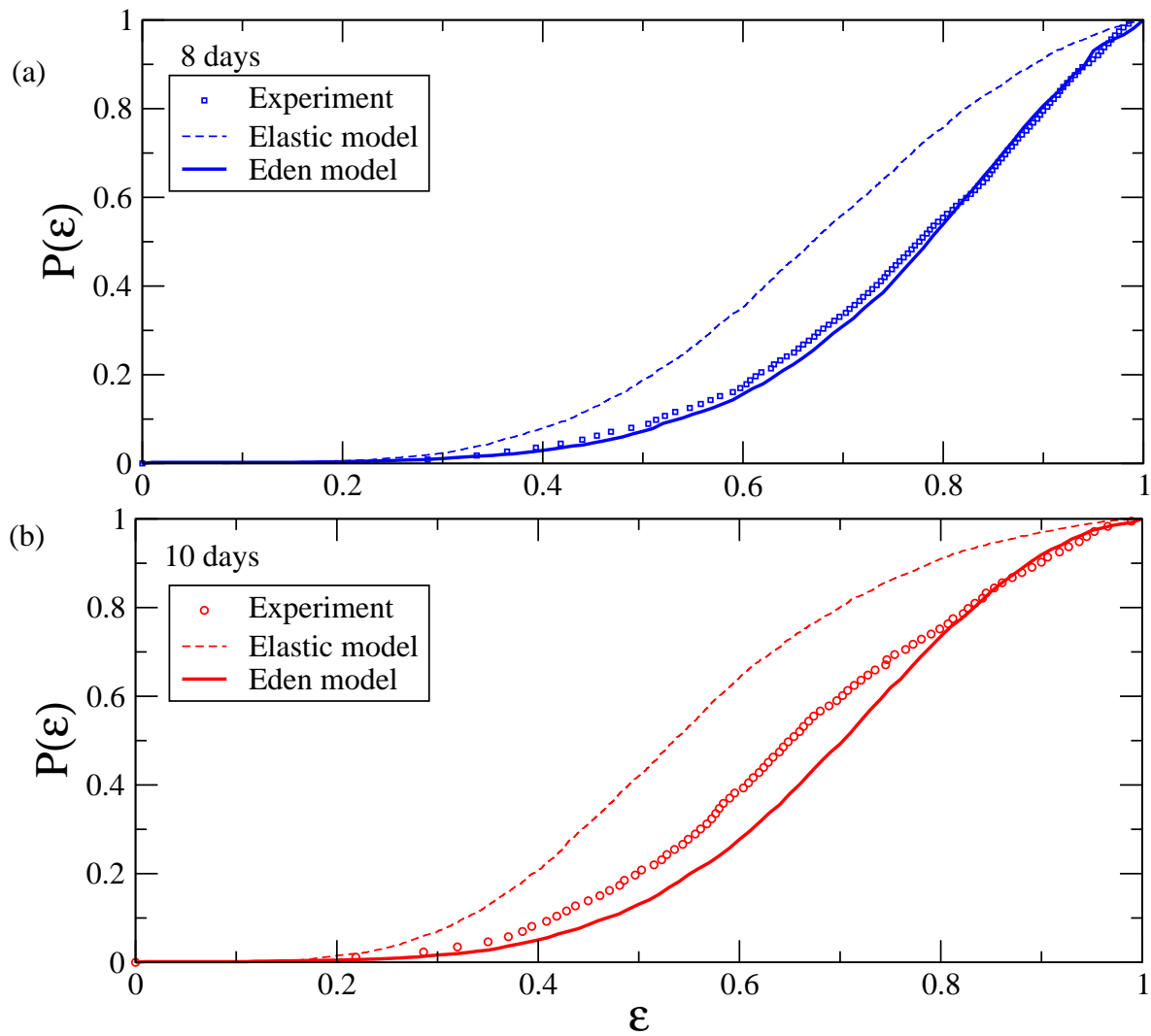}
\caption{The distribution of eccentricities for colonies grown for (a) 8 days and (b) 10 days. The dashed line is
the result of the cell mechanics model, the solid line is the result of the Eden model. In both cases, the models
have been simulated with the parameters obtained from the fit of the colony size distribution with the continuum time branching process.
}
\label{fig:pecc}
\end{figure}

Cancer cell colonies come in different sizes but also in different shapes.
A natural measure of the shape of a colony is the inertia moment tensor,
which we define with respect to the centre of mass of each colony when 
at each pixel is assigned unit mass.
The principal moments of inertia $\lambda_{M}$ and $\lambda_{m}$, where $M$ denote the maximum and $m$ denote the minimum eigenvalue, and the corresponding principal axes  found by diagonalizing the inertia moment tensor. 
Here we characterize the shape of each cluster with the eccentricity defined as
\begin{equation}
\epsilon \equiv \sqrt{1-\left(\frac{\lambda_{m}}{\lambda_{M}}\right)^2}.
\end{equation}
In Fig. \ref{fig:pecc} we report the cumulative distribution of eccentricities $P(\epsilon)$ 
for colonies grown 8 or 10 days. The figure shows that the distribution shifts to
the left with time, suggesting that the colonies become more isotropic as they grow. 
In percolation clusters, the ratio $\lambda_M/\lambda_m$ is found to
display universal corrections to scaling \cite{family1985}. Similar relations have been obtained
in other models, such as polymers \cite{aronovitz1986} and lattice animals \cite{aronovitz1987}.
It would be interesting to compute directly the correction to scaling exponent for our clusters
in analogy with Ref.  \cite{family1985}, but our data appear to be insufficient for this purpose.

\subsection{Modeling shape fluctuations: Elastic cell model}
The continuous time branching process used to describe the colony size distribution does not consider
geometry and does not provide any information on the shape of the colony. To overcome this limitation, we
consider a model of elastic spherical cells dividing randomly in two dimensions. The cells interact with 
a simple Hertz pair potential \cite{drasdo1995}
\begin{equation}
V(r_{ij}) = \frac{4E (2R-r_{ij})}{15(1-\nu^2)} \sqrt{R \over 2}, \mbox{ for } r_{ij} <2R
\end{equation}
where $R$ is the cell radius, $r_{ij}$ is the distance between the cell centers, $E$ is the Young modulus and $\nu$ is the
Poisson ratio. Cell division occurs randomly at rate $\gamma$. To this end we use the Gillespie algorithm \cite{gillespie1976}:
we randomly select a cell and perform a cell division at time sequences where the step at time $t$  is chosen according to a Poisson distribution with rate $s \gamma$ where $s$ 
is the number of cells in the colony at time $t$. Choosing $\gamma=0.55$ divisions/day, we reproduce the experimentally 
measured size distribution. Cell division is simulated by replacing the dividing cell by two randomly oriented 
new cells placed at distance $r=R/10$. The system is then relaxed to mechanical equilibrium, locally 
minimizing the total elastic energy
\begin{equation}
E_{el} = \sum_{\langle ij \rangle} V(r_{ij}),
\end{equation}
where the sum is restricted to cells in contact.

Using this model we simulate a set of 4000 colonies after 8 and 10 days of growth. The resulting cumulative
distribution of eccentricities is reported in Fig. \ref{fig:pecc}. We can see systematic differences with respect
to the experimental data. 

\subsection{Modeling shape fluctuations: Eden model}
A possible explanation for this result is that cell division does not occur homogeneously
throughout the colony as we assume in the model. Experimental results for other tumors show that cell division is
mostly confined to the outer rim of the colony, with little division occurring in the bulk \cite{bru2003}. We note that
perimeter growth is somewhat incompatible with our branching process analysis of the size distribution, since
asymptotically the growth of the colony would not be exponential. For the short times and small cluster sizes analyzed here, however,
growth is indeed exponential as shown by experiments \cite{bru2003} and models \cite{drasdo2012}, so that
a mean-field analysis based on branching processes yields reliable results.

To take into account the fact that the rate of  division depends on the cell location we have to go beyond
a mean-field description.
The simplest growth model to describe a colony in which cell division is restricted to the periphery is
the Eden model \cite{eden1961}, originally devised to describe bacterial colonies, but recently shown to reproduce
the roughness of cancer cell colonies \cite{huergo2012}.  The Eden model can be simulated on a lattice
as in the original paper \cite{eden1961} by randomly growing sites whose nearest neighbors are already occupied.
In this form, however, Eden clusters inherit the anisotropy of the square lattice \cite{freche1985}. This problem 
is overcome in the off-lattice version of the model in which circular particles are added randomly to the aggregate taking care
to avoid any overlap between the particles \cite{wang1995,ferreira2006,alves2011}. Lattice anisotropy can   
be avoided even in the square lattice by using appropriate growth rules as shown in Ref. \cite{paiva2007}. 
Here use the latter approach to grow Eden clusters.

We consider a two dimensional square lattice and start with an initial seed. At each step we select an occupied site in the
lattice with at least one empty nearest neighbor site. We randomly select one of the empty sites $j$ and 
occupy it with a probability  $p$ depending on $n_j$ the the number of occupied nearest neighbors of the site $j$
as $p=n_j/4$. This particular version of the Eden model has been shown to result in clusters
that are asymptotically circular thus reducing lattice anisotropy \cite{paiva2007}. We use this model to generate
a set of clusters with size distributions similar to the experimental ones. This is done in practice implementing
the same Gillespie algorithm employed for the elastic cell model. We next compute the distribution of eccentricities
after 8 and 10 days. The results, shown in Fig. \ref{fig:pecc}, are in good agreement with experimental data.

\section{Discussion}

In this paper we have performed a statistical analysis of the growth of melanoma cell colonies 
using the crystal violet assay, a standard method in cancer biology. 
While most (but not all \cite{kimmel1991}) previous studies of cancer
colony formation focused on the morphology of a single colony \cite{bru2003,buceta2005,huergo2011,huergo2012},
we consider the time evolution of the distribution of sizes and shapes of different colonies. 
The colony size distribution is quantitatively described by a birth and death process, with a negligible rate of cell death.
We next analyze the fluctuations of colony shapes as described by the eccentricity distribution
obtained from the eigenvalues of the moment of inertia tensor. We compare the experimentally measured
distributions with two simple model of cell proliferation: a mechanical model in which cell duplicate and
interact elastically and a geometrical model in which cell duplicate only if the surrounding space is empty.
The latter case correspond to the Eden model \cite{eden1961} which we simulate using a lattice version
that minimize systematic lattice anisotropies \cite{paiva2007}. 

Our results indicate that the Eden model does a better job at reproducing the experiments than the elastic
particle mode, indicates that cell division is hindered just by geometrical constraints, since there is no difference 
in terms of nutrients or oxygen between the interior and the boundary of the colonies here. This is confirmed by experimental results
showing that duplication is reduced in the interior of the colony \cite{bru2003}. A purely elastic model
as the one we employ here is not adequate to describe the data, but one could introduce geometrical constraints
in this type of models by reducing the duplication rate for cells under compression \cite{drasdo2012}.

We notice that the boundary surface of the Eden model clusters is self-affine with a roughness exponent 
described by the Kardar-Parisi-Zhang (KPZ) equation \cite{kardar1986}. Other experiments 
on individual colony growth indicate that the scaling behavior of the surface falls into the KPZ universality class\cite{huergo2011,huergo2012}.
Earlier results suggesting a different universality class \cite{bru2003} have been
attributed to artefacts in the statistical analysis \cite{buceta2005,pastor2007}.  Our colonies are too small to
measure the roughness exponent directly, but it is still interesting to remark that the Eden model
describes their shapes. In conclusions, our results provide an illustration of the wealth of quantitative information that can be extracted by
a standard biological method such as the crystal violet assay.

\section*{Acknowledgements}
CAMLP, JPS and SZ acknowledge the hospitality of the Aspen Center for Physics, which is supported by the National Science Foundation Grant  PHY-1066293. JSP was supported by NCI U54CA143876.

%\bibliographystyle{iopart-num} 
%\bibliography{../stem}

\begin{thebibliography}{10}
\expandafter\ifx\csname url\endcsname\relax
  \def\url#1{{\tt #1}}\fi
\expandafter\ifx\csname urlprefix\endcsname\relax\def\urlprefix{URL }\fi
\providecommand{\eprint}[2][]{\url{#2}}

\bibitem{kimmel1991}
Kimmel M and Axelrod D~E 1991 {\em Journal of Theoretical Biology\/} {\bf 153}
  157 -- 180
  

\bibitem{bru2003}
Bru A, Albertos S, Subiza J~L, Garcia-Asenjio J~L and Bru I 2003 {\em
  Biophysical journal\/} {\bf 85} 2948--2961

\bibitem{buceta2005}
Buceta J and Galeano J 2005 {\em Biophysical Journal\/} {\bf 88} 3734 -- 3736
 
\bibitem{huergo2011}
Huergo M~A~C, Pasquale M~A, Gonz\'alez P~H, Bolz\'an A~E and Arvia A~J 2011
  {\em Phys. Rev. E\/} {\bf 84}(2) 021917


\bibitem{huergo2012}
Huergo M~A~C, Pasquale M~A, Gonz\'alez P~H, Bolz\'an A~E and Arvia A~J 2012
  {\em Phys. Rev. E\/} {\bf 85}(1) 011918


\bibitem{drasdo1995}
Drasdo D, Kree R and McCaskill J 1995 {\em Physical Review E\/} {\bf 52}
  6635--6657
 

\bibitem{galle2005}
Galle J, Loeffler M and Drasdo D 2005 {\em Biophysical Journal\/} {\bf 88} 62
  -- 75
  

\bibitem{galle2009}
Galle J, Hoffmann M and Aust G 2009 {\em Journal of Mathematical Biology\/}
  {\bf 58}(1) 261--283

\bibitem{drasdo2012}
Drasdo D and Hoehme S 2012 {\em New Journal of Physics\/} {\bf 14} 055025

\bibitem{loeffler1987}
Loeffler M, Potten C and Wichmann H 1987 {\em Virchows Archiv Abteilung B Cell
  Pathology\/} {\bf 53} 286--300

\bibitem{block2007}
Block M, Schoell E and Drasdo D 2007 {\em Physical Review Letters\/} {\bf 99}

\bibitem{chatelain2011}
Chatelain C, Balois T, Ciarletta P and Amar M~B 2011 {\em New Journal of
  Physics\/} {\bf 13} 115013

\bibitem{harris89}
Harris T~E 1989 {\em The theory of branching processes\/} (Dover, New York)

\bibitem{kimmel2002}
Kimmel M and Axelrod D~E 2002 {\em Branching processes in biology\/} (Springer,
  New York)

\bibitem{vogel1968}
Vogel H, Niewisch H and Matioli G 1968 {\em J Cell Physiol\/} {\bf 72} 221--8

\bibitem{matioli1970}
Matioli G, Niewisch H and Vogel H 1970 {\em Rev Eur Etud Clin Biol\/} {\bf 15}
  20--2

\bibitem{potten1988}
Potten C~S and Morris R~J 1988 {\em J Cell Sci Suppl\/} {\bf 10} 45--62

\bibitem{clayton2007}
Clayton E, Doupé D~P, Klein A~M, Winton D~J, Simons B~D and Jones P~H 2007
  {\em Nature\/} {\bf 446} 185--9

\bibitem{antal2010}
Antal T and Krapivsky P~L 2010 {\em Journal of Statistical Mechanics: Theory
  and Experiment\/} {\bf 2010} P07028

\bibitem{itzkovitz2012}
Itzkovitz S, Blat I~C, Jacks T, Clevers H and van Oudenaarden A 2012 {\em
  Cell\/} {\bf 148} 608--19

\bibitem{michor2005}
Michor F, Hughes T~P, Iwasa Y, Branford S, Shah N~P, Sawyers C~L and Nowak M~A
  2005 {\em Nature\/} {\bf 435} 1267--70

\bibitem{ashkenazi2008}
Ashkenazi R, Gentry S~N and Jackson T~L 2008 {\em Neoplasia\/} {\bf 10}
  1170--1182

\bibitem{michor2008}
Michor F 2008 {\em J Clin Oncol\/} {\bf 26} 2854--61

\bibitem{tomasetti2010}
Tomasetti C and Levy D 2010 {\em Proc Natl Acad Sci U S A\/} {\bf 107}
  16766--71

\bibitem{laporta2012}
La~Porta C~A~M, Zapperi S and Sethna J~P 2012 {\em PLoS Comput Biol\/} {\bf 8}
  e1002316

\bibitem{eden1961}
Eden M 1961 {\em Proceedings of the Fourth Berkeley Symposium on Mathematical
  Statistics and Probability\/} vol~IV ed Neyman F (Univ. of California Press,
  Berkeley, 1961) p 233

\bibitem{taghizadeh2010}
Taghizadeh R, Noh M, Huh Y~H, Ciusani E, Sigalotti L, Maio M, Arosio B, Nicotra
  M~R, Natali P, Sherley J~L and La~Porta C~A~M 2010 {\em PLoS One\/} {\bf 5}
  e15183

\bibitem{marr1980}
Marr D and Hildreth E 1980 {\em Proceedings of the Royal Society of London.
  Series B. Biological Sciences\/} {\bf 207} 187--217

\bibitem{hoshen76}
Hoshen J and Kopelman R 1976 {\em Phys. Rev. B\/} {\bf 14}(8) 3438--3445

\bibitem{family1985}
Family F, Vicsek T and Meakin P 1985 {\em Phys. Rev. Lett.\/} {\bf 55}(7)
  641--644 

\bibitem{aronovitz1986}
Aronovitz J~A and Nelson D~R 1986 {\em J. Phys. France\/} {\bf 47} 1445--1456
  
\bibitem{aronovitz1987}
Aronovitz J~A and Stephen M~J 1987 {\em Journal of Physics A: Mathematical and
  General\/} {\bf 20} 2539

\bibitem{gillespie1976}
Gillespie D~T 1976 {\em Journal of Computational Physics\/} {\bf 22} 403 -- 434

\bibitem{freche1985}
Freche P, Stauffer D and Stanley H~E 1985 {\em Journal of Physics A:
  Mathematical and General\/} {\bf 18} L1163

\bibitem{wang1995}
Wang C~Y, Liu P~L and Bassingthwaighte J~B 1995 {\em Journal of Physics A:
  Mathematical and General\/} {\bf 28} 2141

\bibitem{ferreira2006}
Ferreira S~C and Alves S~G 2006 {\em Journal of Statistical Mechanics: Theory and
  Experiment\/} {\bf 2006} P11007

\bibitem{alves2011}
Alves S~G, Oliveira T~J and Ferreira S~C 2011 {\em EPL (Europhysics Letters)\/}
  {\bf 96} 48003

\bibitem{paiva2007}
Paiva L~R and Ferreira S~C 2007 {\em Journal of Physics A: Mathematical and
  Theoretical\/} {\bf 40} F43

\bibitem{kardar1986}
Kardar M, Parisi G and Zhang Y~C 1986 {\em Phys. Rev. Lett.\/} {\bf 56}(9)
  889--892 

\bibitem{pastor2007}
Pastor J and Galeano J 2007 {\em Central European Journal of Physics\/} {\bf
  5}(4) 539--548 
\end{thebibliography}

\providecommand{\newblock}{}

\newpage

\end{document}